\RequirePackage{amsmath}
\documentclass{llncs}

\usepackage{amsmath}
\usepackage{amssymb}
\usepackage{array}
\usepackage{booktabs}
\usepackage{color}
\usepackage{float}
\usepackage[T1]{fontenc}
\usepackage{graphicx}
\usepackage{ifpdf}
\usepackage[utf8]{inputenc}
\usepackage{keyval}
\usepackage{listings}
\usepackage{lmodern}
\usepackage{moresize}
\usepackage{multirow}
\usepackage[numbers,sort&compress,square,sectionbib]{natbib}
\usepackage{paralist}
\usepackage{rotating}
\usepackage{silence}
\usepackage{soul}
\usepackage{srcltx}
\usepackage{subfig}
\usepackage{textcomp}
\usepackage[normalem]{ulem}
\usepackage{url}
\usepackage{wrapfig}
\usepackage{xcolor}
\usepackage{xspace}

\definecolor{listinggray}{gray}{0.95}
\definecolor{darkgray}{gray}{0.7}
\definecolor{commentgreen}{rgb}{0, 0.4, 0}
\definecolor{darkblue}{rgb}{0, 0, 0.6}
\definecolor{purple}{rgb}{0.6, 0, 0.6}
\definecolor{middleblue}{rgb}{0, 0, 0.75}
\definecolor{darkred}{rgb}{0.4, 0, 0}
\definecolor{brown}{rgb}{0.5, 0.5, 0}
\definecolor{dkgreen}{rgb}{0,0.5,0}
\definecolor{orange}{rgb}{1,.5,0}
\definecolor{dandelion}{cmyk}{0,0.29,0.84,0}

\makeatletter
\def\cyanuwave{\bgroup \markoverwith{\lower3.5\p@\hbox{\sixly \textcolor{cyan}{\char58}}}\ULon}
\def\reduwave{\bgroup \markoverwith{\lower3.5\p@\hbox{\sixly \textcolor{red}{\char58}}}\ULon}
\def\blueuwave{\bgroup \markoverwith{\lower3.5\p@\hbox{\sixly \textcolor{blue}{\char58}}}\ULon}
\font\sixly=lasy6 
\makeatother

\newif\ifdraft{}
\ifdraft{}
 \newcommand{\amnote}[1]{\textcolor{blue}   {***AM\@: #1}}
 \newcommand{\jhanote}[1]{\textcolor{red}   {***SJ\@: #1}}
 \newcommand{\mtnote}[1]{\textcolor{orange} {***MT\@: #1}}
 \newcommand{\mmnote}[1]{\textcolor{violet} {***MM\@: #1}}
 \newcommand{\msnote}[1]{\textcolor{magenta}{***MS\@: #1}}
\else
 \newcommand{\msnote}[1]{}
 \newcommand{\amnote}[1]{}
 \newcommand{\jhanote}[1]{}
 \newcommand{\mmnote}[1]{}
 \newcommand{\mtnote}[1]{}
\fi

\newcommand{\bw}{\I{Blue\,Waters}}

\newcommand{\titan}{\I{Titan}}

\newcommand{\I}[1]{\textit{#1}\xspace}
\newcommand{\B}[1]{\textbf{#1}\xspace}
\newcommand{\T}[1]{\texttt{#1}\xspace}

\DeclareFixedFont{\ttb}{T1}{txtt}{bx}{n}{9} 
\DeclareFixedFont{\ttm}{T1}{txtt}{m}{n}{9}  
\definecolor{deepblue}{rgb}{0,0,0.5}
\definecolor{deepred}{rgb}{0.6,0,0}
\definecolor{deepgreen}{rgb}{0,0.5,0}

\newcommand{\pythonstyle}{
  \lstset{
    language=Python,
    backgroundcolor=\color{white},
    basicstyle=\footnotesize\ttfamily,
    basewidth  = {0.6em,0.6em},
    breaklines=true,
    otherkeywords={self},
    keywordstyle=\ttb\color{deepblue},
    emph={MyClass,__init__},
    emphstyle=\ttb\color{deepred},
    stringstyle=\color{deepgreen},
    commentstyle=\color{red},
    frame=tb,
    showstringspaces=false
  }
}

\lstnewenvironment{python}[1][]{
  \pythonstyle{}
  \lstset{#1}
}{\vspace*{-0.5em}}

\colorlet{punct}{red!60!black}
\definecolor{delim}{RGB}{20,105,176}
\colorlet{numb}{magenta!60!black}

\newcommand{\lstsetjson}{
  \lstset{
    language=json,
    backgroundcolor=\color{white},
    basicstyle=\footnotesize\ttfamily,
    basewidth={0.6em,0.6em},
    breaklines=true,
    keywordstyle=\ttb\color{deepblue},
    frame=tb,
    showstringspaces=false
    stringstyle=\ttfamily\color{BurntOrange},
    showstringspaces=false,
    tabsize=2,
  }
}

\lstdefinelanguage{json}{
    morekeywords={bw_aprun,
                  description,
                  notes,
                  workdir,
                  valid_roots,
                  virtenv,
                  rp_version,
                  virtenv_mode,
                  schemas,
                  gsissh,
                  job_mgr_url,
                  file_mgr_url,
                  stage_cacerts,
                  default_queue,
                  lrms,
                  agent_type,
                  agent_scheduler,
                  agent_spawner,
                  agent_launch_method,
                  task_launch_method,
                  mpi_launch_method,
                  pre_bootstrap_1},
    breaklines=true,
    literate=
      {:}{{{\color{punct}{:}}}}{1}
      {,}{{{\color{punct}{,}}}}{1}
      {\{}{{{\color{delim}{\{}}}}{1}
      {\}}{{{\color{delim}{\}}}}}{1}
      {[}{{{\color{delim}{[}}}}{1}
      {]}{{{\color{delim}{]}}}}{1},
}

\lstnewenvironment{json}[1][]{
  \lstsetjson{}
  \lstset{#1}
}{\vspace*{-0.5em}}

\begin{document}

\title{Using Pilot Systems to Execute Many Task Workloads on Supercomputers}

\author{Andre Merzky\inst{1} 
        \and Matteo Turilli\inst{1} 
        \and Manuel Maldonado\inst{1}
        \and Mark Santcroos\inst{1}
        \and Shantenu Jha\inst{1,2}}
\institute{RADICAL Laboratory, Electrical and Computer Engineering, Rutgers
University, Piscataway, NJ, USA 
        \and Brookhaven National Laboratory, Upton, NY, USA}

\maketitle

\begin{abstract}
High performance computing systems have historically been designed to support applications comprised of mostly monolithic, single-job workloads. Pilot systems decouple workload specification, resource selection, and task execution via job placeholders and late-binding. Pilot systems help to satisfy the resource requirements of workloads comprised of multiple tasks. RADICAL-Pilot (RP) is a modular and extensible Python-based pilot system. In this paper we describe RP's design, architecture and implementation, and characterize its performance. RP is capable of spawning more than 100 tasks/second and supports the steady-state execution of up to 16K concurrent tasks. RP can be used stand-alone, as well as integrated with other application-level tools as a runtime system.

\end{abstract}

\section{Introduction}\label{sec:intro}

Traditionally, advances in high-performance scientific computing have focused
on the scale, performance and optimization of a workload with a large but
single task, and less on workloads comprised of multiple tasks.
High-performance workflows and scalable computation of ensemble workloads are
becoming increasingly important and are highly relevant to exploit post-Moore
parallelism. As a result, the number and type of applications that can be
formulated as workflows or ensembles is vast and span many scientific
domains.

Applications with workloads comprised of multiple tasks impose sophisticated
execution and advanced resource management
requirements~\cite{better-resource}. High-performance computing (HPC) systems
have been designed to support applications comprised of mostly monolithic,
single-job workloads. For example, HPC systems have been designed and
operated to maximize overall system utilization, which typically entails
static resource partitioning across jobs and users. Thus, there is a tension
between the resource requirements of workloads comprised of many tasks, and
the capabilities of the traditional HPC resource management as well as their
usage policies. This tension motivates middleware that can efficiently manage
the ability to support the resource requirements of many task workloads
without compromising traditional capabilities of HPC systems.

Enter pilot systems. The authors in Ref.~\cite{review_pilotreview} defined
the properties of the Pilot paradigm, and its relevance in the execution of
workloads comprised of multiple tasks. A defining element of the Pilot
paradigm is the execution of a workload via multi-entity and multi-stage
scheduling on resource placeholders. Systems implementing the Pilot paradigm
submit job placeholders (i.e., pilots) to the scheduler of resources. Once
active, each pilot accepts and executes tasks directly submitted to it by the
application. In this way, pilot systems decouple workload specification,
resource selection, and task execution via job placeholders and late-binding.

Pilot systems address two apparently contradictory requirements: accessing
HPC resources via their centralized schedulers, and letting applications
independently schedule tasks on the acquired portion of resources. Thus,
pilot systems provide a simple solution to the rigid resource management
model historically found in HPC systems. Not surprisingly, many workflow
management systems use pilot systems. Surprisingly, in spite of the
acceptance and uptake of pilot systems, to the best of our knowledge, there
are no general purpose implementations capable of working in production with
multiple HPC resources, including leadership class machines.

In this paper, we discuss the design, architecture and implementation of
RADICAL-Pilot (RP) (\S3). RP is a pilot system that fully implements the
concepts and capabilities of the Pilot paradigm.  The implementation of RP
differs from other pilot systems mostly in terms of API, portability, and
introspection. Implemented in Python, RP is a self-contained pilot system
which can be used to provide a runtime system for workloads comprised of
multiple tasks. In \S4, we discuss how RP provides pilot capabilities on Cray
systems such as \bw{} and \titan{}.  We experimentally characterize RP at
multiple levels in \S5: we study the performance of individual components of
RP, followed by the integrated performance of its Agent. We then investigate
the resource utilization and performance of both the native and enhanced
scheduling algorithms.

The absolute performance of the enhanced scheduler is less important than the
ability to enhance performance of the scheduler via extensions and customized
scheduling algorithms. This reiterates the core contribution of this paper: a
careful description of the design and implementation of RP, highlighting its
use of multi-level and multi-entity scheduling.

\section{Related Work}\label{sec:related}

Traditionally, HPC systems such as Crays have been designed to best support
monolithic workloads. However, the workload of many important scientific
applications is constructed out of spatially and temporally heterogeneous
tasks that are often dynamically inter-related, where those tasks require
compute, memory and communication capabilities exceeding what single node
machine can provide, and where the \I{overall} workload requirements are
comparable to or exceeding those of classic HPC
workloads~\cite{preto2014fast,cheatham2015impact,sugita1999replica}. These
workloads can benefit from being executed at scale on supercomputers (e.g.,
\bw{} and \titan{}, both Cray systems), but a tension exists among the
workloads' resource utilization requirements like rapidly and repeatedly
acquiring a certain amount of cores over time, the capabilities of the HPC
system software, and their usage policies. Pilot systems have the potential
to relieve this tension but their adoption for this class of HPC systems
present several challenges that, so far, have not been fully addressed.

Since 1995, more than twenty pilot systems have been
developed~\cite{review_pilotreview}. Most of these systems are tailored to
specific workloads, resources, interfaces, or development models. Most pilot
systems have been implemented to optimize the throughput of single-core (or
single-node), short-lived, uncoupled tasks~\cite{review_pilotreview}. Some
notable examples are: HTCondor with Glidein on OSG~\cite{pordes2007open}, one
of the most widely used pilot systems for the execution of mostly single-core
workloads; the pilot systems developed for the LHC communities which execute
millions of jobs a week~\cite{maeno2014evolution} and are specialized in
supporting LHC workloads on specific resources like those of WLCG\@; the
light-weight execution framework called Falkon, which represents an early
stand-alone pilot system for HPC environment~\cite{raicu2007falkon}; and
Coasters, developed mostly to support the Swift workflow
system~\cite{wilde2011swift}.

One of the major challenges in developing a general-purpose pilot system,
capable of executing multi-task workloads on supercomputers, is supporting
multiple task launch methods, each with a specific set of limitations. For
example, Cluster Compatibility Mode (CCM)~\cite{CCM} is designed to provide
services analogous to those of Beowulf clusters but is not generally
available on all Cray installations and, when present, access to it varies
per system. The Application Level Placement Scheduler
(ALPS)~\cite{karo2006application} system, provides launch functionality for
running executables on compute nodes but limits the number of concurrent
applications a user can run by default. The Open Run-Time
Environment~\cite{castain:orte2007}, a component of the OpenMPI MPI
implementation, supports distributed high-performance computing applications
operating in a heterogeneous environment but the degree of adoption and
support varies across Cray systems.

Tools have been developed to support spatially and/or temporally
heterogeneous tasks on Crays but many of these tools are built on top of CCM,
ALPS, or use single MPI allocations. As such, they are not able to support
task heterogeneity or reach the necessary level of execution concurrency. For
example, TaskFarmer~\cite{taskfarmer}, a tool developed at LBNL, enables the
user to execute a list of system commands from a task file, allowing
single-core or single-node tasks to be run within a single \T{mpirun}
allocation. Wraprun~\cite{wraprun}, a utility developed at ORNL, enables
independent execution of multiple MPI applications under a single \T{aprun}
call. QDO~\cite{qdo}, a lightweight high-throughput queuing system for
workflows that have many small tasks has to use the resource batch system for
job submission. MySGE~\cite{canon2012my}, another tool developed at LBNL that
allows users to create a private Sun GridEngine cluster on large parallel
systems, but is only available on NERSC resources. Python Task Farm
(PTF)~\cite{pythontaskfarm}, a utility for running serial Python programs as
multiple independent copies of a program over many cores, is available only
on ARCHER (at EPCC).

The Pilot paradigm has proven sufficiently useful that resource management
systems have begun to include pilot capabilities either as separate tooling,
or as part of their implementation. For example, Flux~\cite{ahn2014flux} is
described as a next-generation Scalable Resource and Job Management Software
(RJMS) for HPC centers that focuses on a new paradigm of resource and job
management. Within this new paradigm, Flux allows resource allocation to be
dynamic (i.e., dynamic workloads), a key design principle of the Pilot
paradigm~\cite{review_pilotreview}. This results in jobs having the ability
to scale up to a maximum requested resources (e.g., CPU cores, GPUs, etc.)
during execution, or to execute workloads (i.e., workloads with different
resource requirements) on a single ``dynamic'' allocation. Unfortunately,
Flux is limited only to the HPC resources that use it as their RJMS\@.
Further, as of the writing of this paper, Flux is still on an Alpha release.

\section{RADICAL-Pilot}\label{sec:rp}

RADICAL-Pilot (RP) is a scalable and interoperable pilot system that
implements the Pilot abstraction to support the execution of diverse
workloads.  We describe the design and architecture of RP, and characterize
the performance of RP's task execution components. These components are
engineered for efficient resource utilization while maintaining the full
generality of the Pilot abstraction.  RP supports several Cray machines,
including \bw{} (NCSA), \titan{} (ORNL), and ARCHER (EPSRC), and a whole
range of other platforms.

\subsection{Overall Architecture}\label{ssec:arch}

\begin{figure}
  \includegraphics[width=\textwidth]{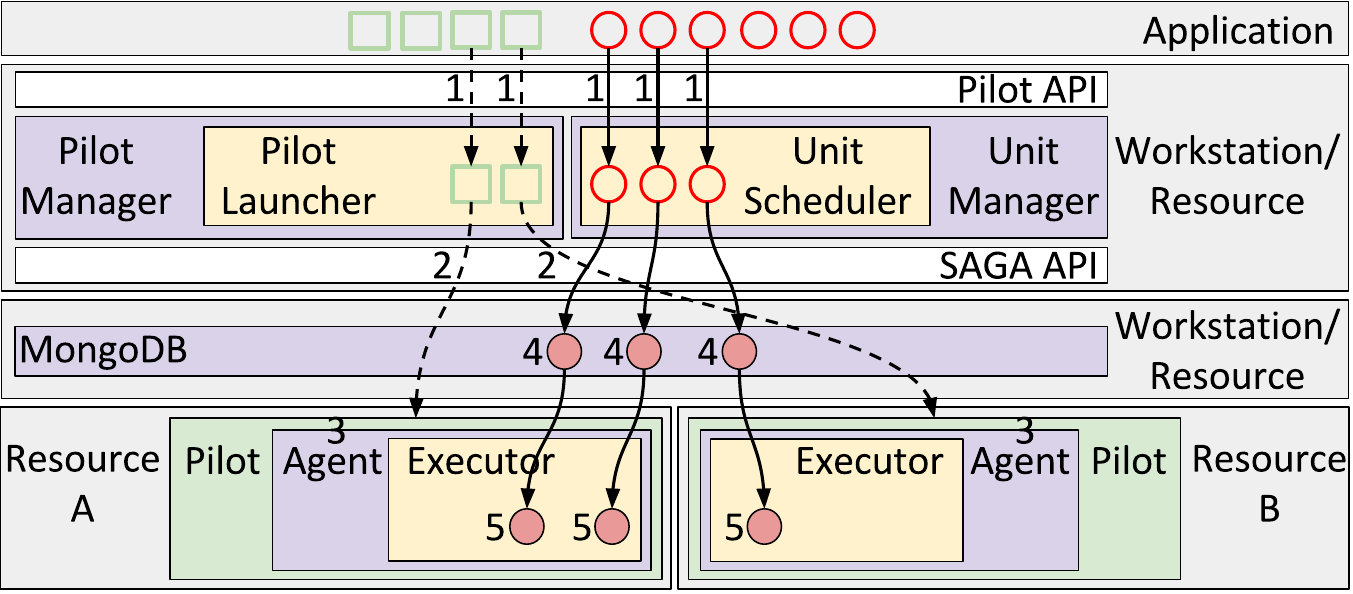}  
  \caption{RADICAL-Pilot Architecture and execution model.}\label{fig:arch}
\end{figure}

RP is a runtime system designed to execute heterogeneous and dynamic
workloads on multiple and diverse resources.  RP's architecture and execution
model are shown in Fig.~\ref{fig:arch}: workloads and pilots are described
via the Pilot-API and passed~\I{(1)} to the RP runtime system, which submits
the pilots, launches the pilots' Agent, and executes the tasks of the
workload on one or more Agents.  RP represents pilots as aggregates of
resources, independent from the architecture and topology of the target
machines, and workloads as a set of units to be executed on the resources of
the pilot. Both pilots and units are stateful entities, each with a
well-defined state model and life cycle.  Their states and state transitions
are managed via the three modules of the RP architecture: PilotManager,
UnitManager, and Agent.

The PilotManager submits pilots to resources via the RADICAL-SAGA
API~\I{(2)}. The SAGA API~\cite{saga-x} implements an adapter for each type
of supported resource, exposing uniform methods for job and data management.
The UnitManager schedules units to pilots' Agent for execution. A MongoDB
database is used to communicate the scheduled workload~\I{(4)} between the
UnitManager and one or more Agent. For this reason, the database instance
needs to be accessible both from the user's workstation and the target
resources, via ssh tunnels that RP creates at runtime, where needed and when
possible.  Each Agent bootstraps on a remote resource, pulls units from the
MongoDB instance, and manages their execution on the cores held by the
pilot~\I{(5)}.

The modules of RP are distributed between the user workstation and the target
resources. The PilotManager and UnitManager are executed on the user
workstation while each Agent runs on the target resources. RP requires Linux
or OS X with Python 2.7 on the workstation but the Agent can execute
different types of units on resources with diverse architectures and software
environments.

\subsection{Programming Model}\label{sec:progmodel}

RP is engineered as a Python library that enables the declarative definition
of resource requirements, and of workloads to execute on them. RP exposes a
pilot-specific application programming interface called Pilot-API and enables
programming of application-specific relationships between resources and
workload in generic Python. In the following code snippets, we walk the
reader through a minimal but complete example of running a workload on \bw{}
using RP\@.

\begin{wrapfigure}{R}{0.5\textwidth}
  \includegraphics[trim=0 0 0 0,clip,width=0.49\textwidth]{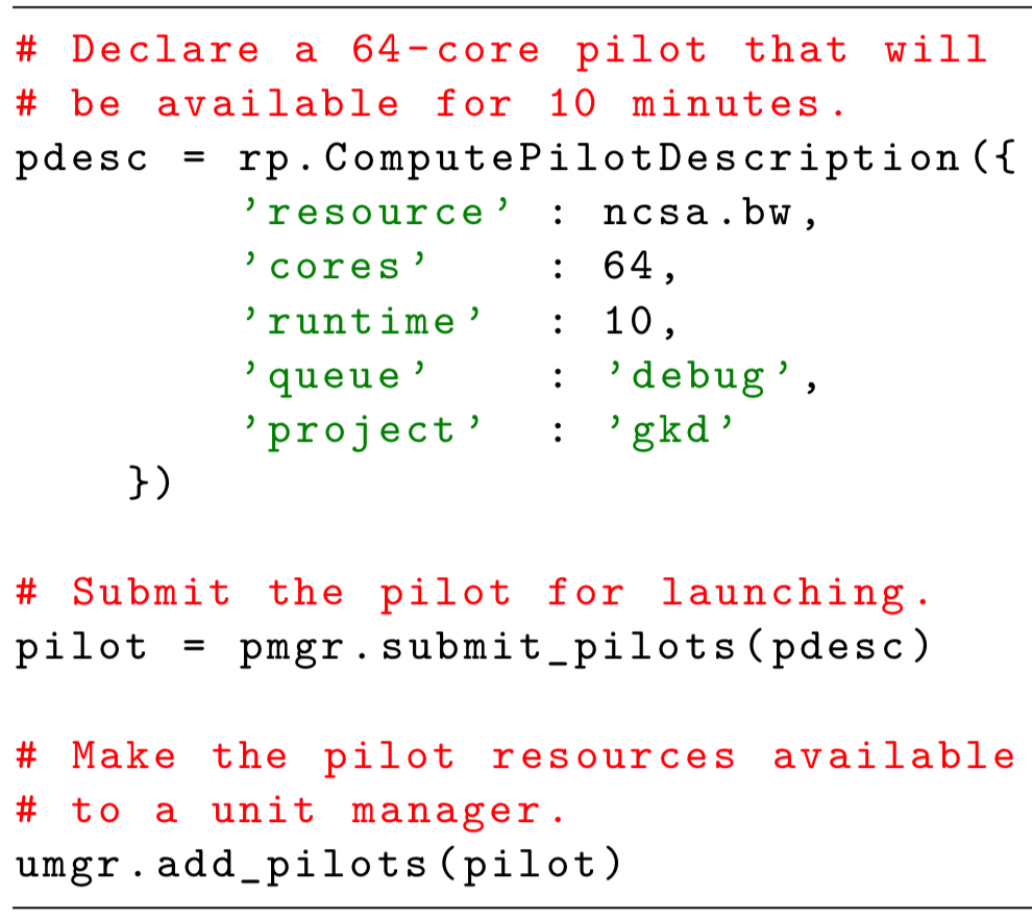}
  \caption{Pilot API\@. Declaration of PilotManager and UnitManager within a
  Session.}\label{lst:managers}
\end{wrapfigure}

In Listing~\ref{lst:managers}, we show the code used to declare the
respective managers for pilots and units, whose lifetime is managed by a
session object. As such, closing a session destroys all its managers.

\begin{figure}
  \subfloat[][]{
    {\includegraphics[width=0.45\columnwidth]{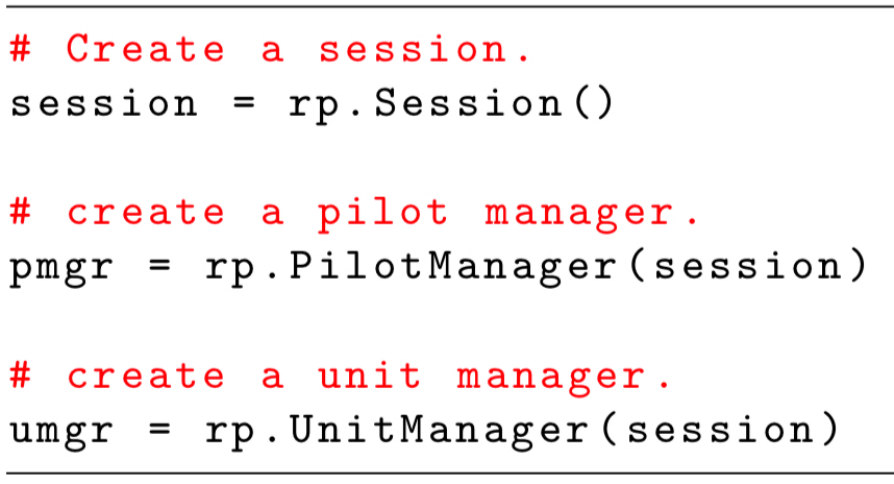}}
  }
  \qquad
  \subfloat[][]{
    {\includegraphics[width=0.45\columnwidth]{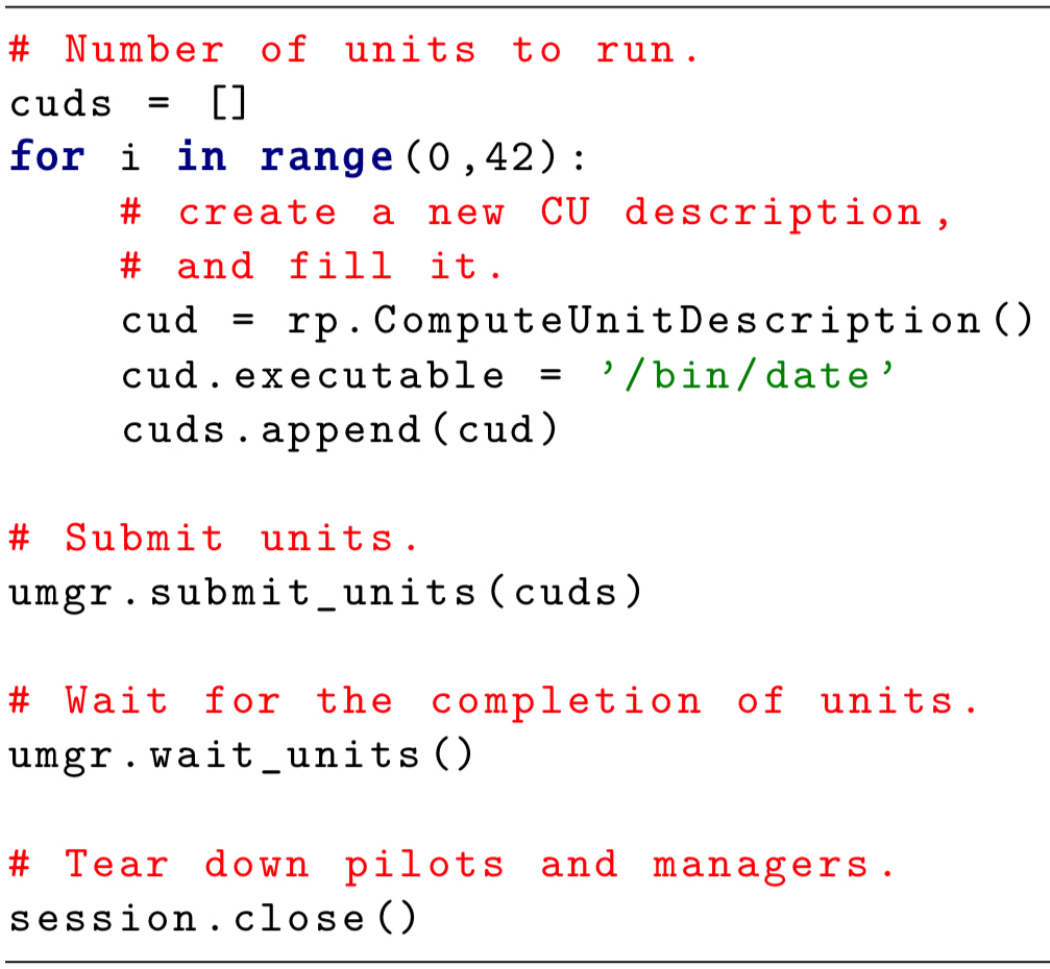}}
  }
  \caption{Pilot API\@. \B{(b)} Declaration of a pilot,
  its subsequent submission to the PilotManager and the association to a
  UnitManager. \B{(c)} Declaration and submission of compute units
  (CU).}\label{lst:pilots_units}
\end{figure}

In Listing~\ref{lst:pilots_units}(a), we declare a pilot
(\T{rp.ComputePilotDescription()}) by specifying the resource on which
it should be instantiated, how many cores it should use, its runtime and,
optionally, to what queue it should be submitted and to what project it
should be charged. Once submitted via the PilotManager
(\T{pmgr.submit\_pilots()}), the pilot is asynchronously queued to the
batch system of the indicated resource. Finally, the pilot is
associated with a UnitManager (\T{umgr.add\_pilots()}) to enable the
execution of units on that pilot.

Finally, in Listing~\ref{lst:pilots_units}(b) we declare a workload by
creating a set of compute units (\T{cuds}) that specify what payload should
be run (\T{/bin/date}). Once created, the compute units are submitted to the
UnitManager (\T{umgr.submit\_units()}) which schedules the unit to a pilot.
The UnitManager can perform early binding (schedule to any known pilot) or
late binding (delay scheduling until pilots become active).  In either case,
once that pilot does become active, it pulls the scheduled units for
execution. The \T{umgr.wait\_units()} call blocks until all the units have
run to completion. Upon its return, the session is closed
(\T{session.close()}) indicating that the workload execution has completed.

\subsection{State and Execution Models}\label{sub:models}

\begin{figure}
  \subfloat[][]{{\includegraphics[width=0.45\columnwidth]{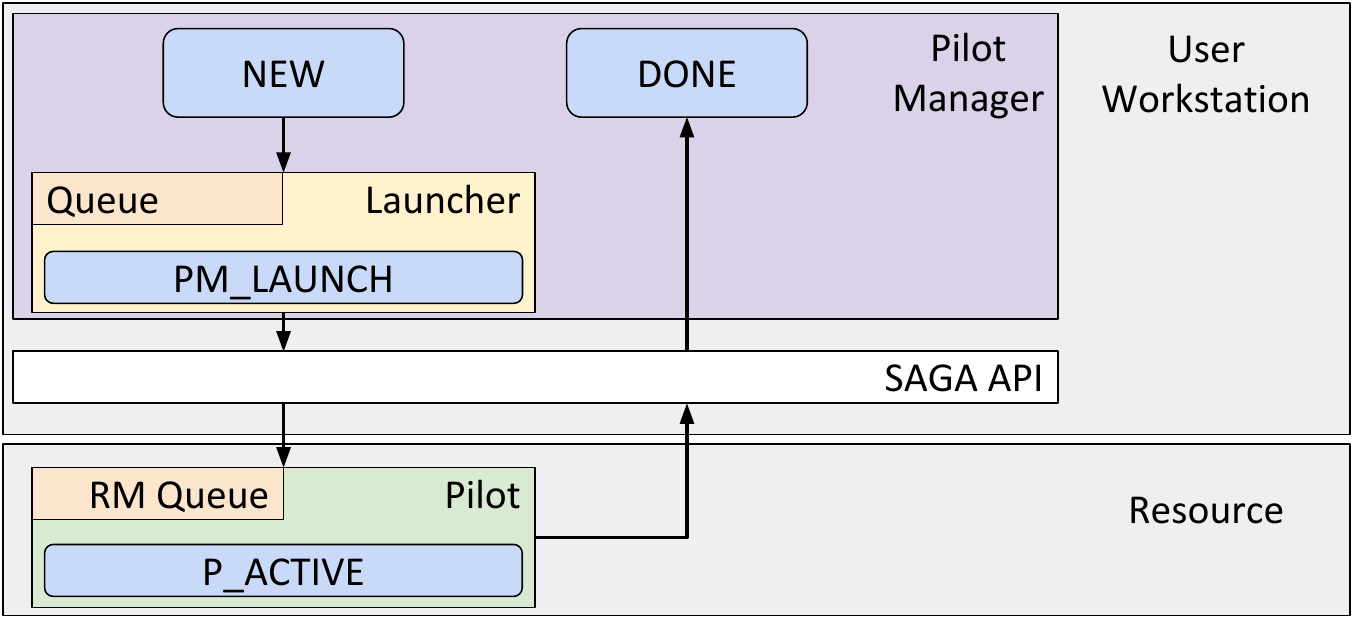}}}
  \qquad
  \subfloat[][]{{\includegraphics[width=0.45\columnwidth]{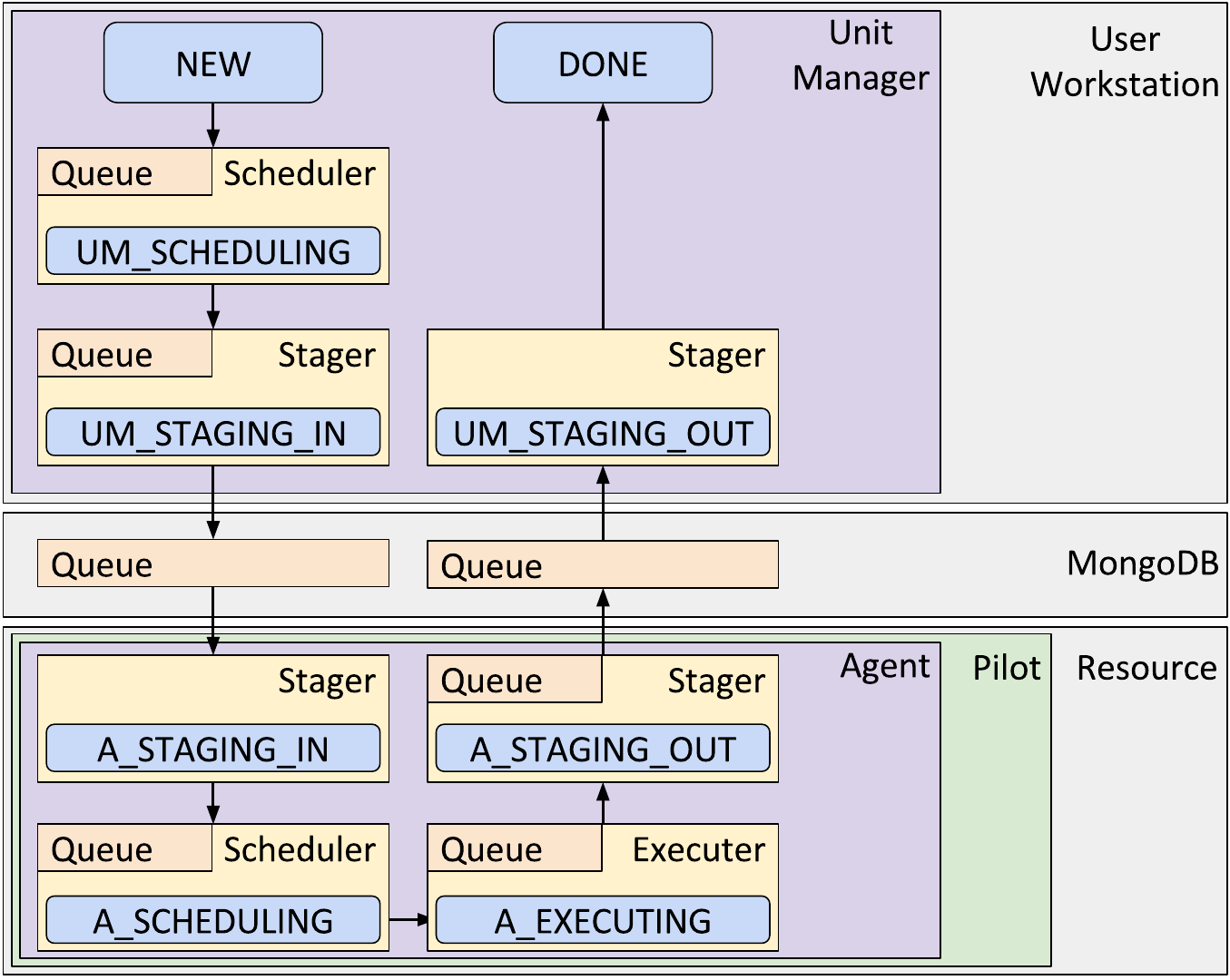}}}
\caption{\B{(a)} Architecture of RP Client and pilot state model.
  \B{(b)} Architecture of RP Agent and unit state
  model.}\label{fig:agent-arch_unit_pilot-state-models}
\end{figure}

The lifespan of pilots has 4 states distributed among the PilotManager,
resource, and pilot instance
(Fig.~\ref{fig:agent-arch_unit_pilot-state-models}a). Pilots are instantiated
in the state \T{NEW} by the PilotManager, wait in a queue to be launched, and
transition to \T{PM\_LAUNCH} when submitted to a Resource Manager (RM) via
the SAGA API\@. Pilots wait in the queue of the RM and, once scheduled,
become \T{P\_ACTIVE}. They remain in this state until the end of their
lifetime, when they transition to \T{DONE}.

The unit state model has 9 states distributed across the UnitManager, MongoDB
instance, and Agent
(Fig.~\ref{fig:agent-arch_unit_pilot-state-models}b). Instantiated in
the state \T{NEW} by the UnitManager, every unit is scheduled on an Agent
(\T{UM\_SCHEDULING}) via a queue on a MongoDB instance. The unit is then
scheduled on the required number of cores held by the Agent's pilot
(\T{A\_SCHEDULING}), and finally executed (\T{A\_EXECUTING}).

When required by a unit, input data are staged in by the UnitManager and
Agent (\T{UM\_STAGING\_IN}, \T{A\_STAGING\_IN}), and output data are staged
out (\T{A\_STAGING\_OUT}, \T{U\_STAGING\_OUT}) to a specified destination,
e.g., local/shared filesystem or user workstation. Both input and output
staging are optional, depending on the requirements of the units. The actual
file transfers are enacted via local OS commands or RADICAL-SAGA, supporting
(gsi)-scp, (gsi)-(s)ftp, and Globus Online.

The state transitions of Fig.~\ref{fig:agent-arch_unit_pilot-state-models}
are sequential, and every transition can fail or be canceled by the
PilotManager or UnitManager. All state transitions are managed by the
PilotManager, UnitManager, and Agent components. The only special case is the
transition of the pilots to the state \T{P\_ACTIVE} which is determined by
the resource's RM and managed by the PilotManager.

\subsection{Agent Architecture}

Depending on the resource architecture, the Agent's Stager, Scheduler, and
Executer components (Fig.~\ref{fig:agent-arch_unit_pilot-state-models}(b))
can be placed on cluster head nodes, machine oriented mini-server (MOM)
nodes, compute nodes, virtual machines, or any combination thereof.  Multiple
Stager and Executer components can coexist in a single Agent, placed on any
service node or compute node of the pilot's resource assignment.  ZeroMQ
communication bridges connect the Agent components, creating a network to
support the transition of units through components. Every unit goes through
the states of Input Staging, Scheduling, Execution \& Output Staging. This
paper investigates different implementations of launch methods, which are
part of the Executer component, responsible for defining and managing the
task execution process.

\section{Enabling RP on Cray systems}\label{sec:rp_cray}

As described in~\cite{cug-2016}, we developed four ways of interfacing RP
with the Cray system software to enable execution of distributed applications
on Cray systems. 

\subsection{Application Level Placement Scheduler (ALPS)}

The ALPS software suite provides launch functionality for running executables
on compute nodes of a Cray system, by interfacing with the \T{aprun}
command. ALPS is the native way to run applications on a Cray from the batch
scheduling system. By default, ALPS limits the user to run up to 1000
applications concurrently within one batch job but in the pilot use-case,
these applications may run only for a very short time. This strains ALPS and
the MOM node, effectively limiting the throughput of concurrent executions to
around 100 applications. Further, ALPS does not allow the user to easily run
more than one task on a single compute node, making it difficult, if not
impossible, to run workloads with tasks requiring single or small amount of
cores and workloads with heterogeneous task size.

\subsection{Cluster Compatibility Mode (CCM)}

Crays are massively parallel processing (MPP) machines and the Cray Compute
Node OS does not provide the full set of Linux services typically found on
Beowulf clusters. CCM is a software suite designed to reduce this gap by
providing services analogous to those of Beowulf clusters when required by
applications. Nonetheless, CCM is not generally available on all Cray
installations and, when present, access to CCM varies per system, requiring
special flags to the job description or submitting to a special queue.

RP hides the CCM deployment differences from the application by operating the
Agent either externally or internally to the CCM cluster created when
submitting a job to the Cray machine. When the Agent runs outside the CCM
cluster, it uses \T{ccmrun} to start tasks. However, this approach still
relies on ALPS, inheriting all the limitations described above. When the
Agent runs within the CCM cluster, only the initial startup of the Agent
relies on ALPS\@. After that, all task launching is done within the cluster,
e.g., by using SSH or MPIRUN, without further interaction with ALPS\@.

\subsection{Open Run-Time Environment (OpenRTE/ORTE)}\label{sub:orte-arch}

The Open Run-Time Environment is a spin-off from the Open-MPI project and is
a critical component of the OpenMPI MPI implementation. It was developed to
support distributed high-performance computing applications operating in a
heterogeneous environment. The system transparently provides support for
interprocess communication, resource discovery and allocation, and process
launch across a variety of platforms. ORTE provides a mechanism similar to
the Pilot concept---it allows the user to create a dynamic virtual machine
(DVM) that spans multiple nodes. In regular OpenMPI usage the lifetime of the
DVM is that of the application, but the DVM can also be made persistent, and
we rely on this particular feature for RP\@. RP supports two different modes
for interacting with the ORTE DVM\@: via orte-submit CLI calls, and via ORTE
library calls. Currently we can not run applications that are linked against
the Cray MPI libraries, but once Cray moves to PMIx\cite{pmix} that issue is
resolved.

\begin{wrapfigure}{R}{0.5\textwidth}
  \includegraphics[trim=0 0 0 0,clip,width=0.49\textwidth]{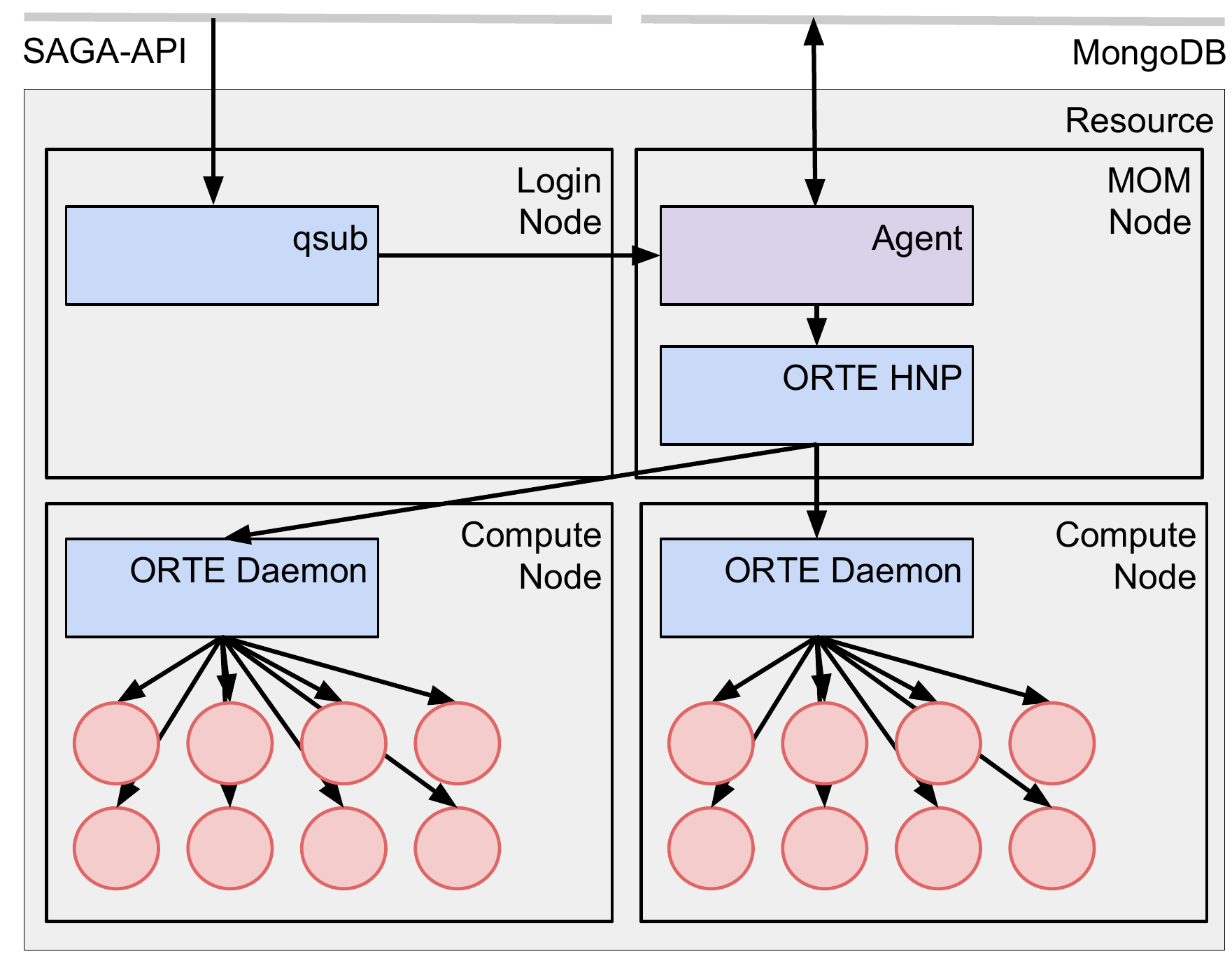}
  \caption{Architecture overview of RP with ORTE
  backend.}\label{fig:orte_arch}
\end{wrapfigure}

Fig.~\ref{fig:orte_arch} shows the layout of the RP agent, the ORTE Head Node
Process (HNP) that manages the DVM on the MOM Node, and the ORTE Daemons that
run on the Compute Nodes.

\subsubsection{Command Line Interface (CLI):}

Recently, ORTE has been extended with tools to expose the creation of the
persistent DVM (\T{orte-dvm}) and the launching of tasks onto that DVM
(\T{orte-submit}). The setup of the DVM requires a single ALPS
interaction, after which all the tasks are executed independent of ALPS\@. As
RP is a Python application and ORTE is implemented in C, we interface the two
systems using the ORTE CLI\@. While this enabled concurrent task execution
and sharing nodes among tasks, we did run into new bottlenecks. The
interaction with the filesystem becomes a limiting factor for task execution
as every task requires the execution of \T{orte-submit}. We also
experience network socket race conditions and system resource limits above
16000 concurrent tasks, as every task requires an \T{orte-submit}
instance that communicates independently with the \T{orte-dvm}. RP has
the ability to spread the execution management of tasks over multiple compute
nodes, addressing the problem of having a large centralized process footprint
for maintaining state about each running process this way.

\subsubsection{C Foreign Function Interface for Python (CFFI):}

CFFI~\cite{cffi} provides a convenient and reliable way to call compiled C
code from Python using interface declarations written in C. 
This mode of operation is similar to the CLI mode, but differs in
the way RP interfaces with ORTE\@: RP launches each task using a library call
instead of the \T{orte-submit} tool. This also allows the re-use of 
network socket, thus further decreasing the per-call overhead. The incentive
for developing this approach was to overcome the limits and overheads imposed
by the CLI approach. We called the resulting launch method ``ORTE-LIB''.

\section{Experiments}\label{sec:exp}

We characterize the performance of the RP Agent by performing experiments to
benchmark individual components and integrated experiments on the Agent as a
whole. The results of experiments on individual components, referred to as
microbenchmarks, characterize the performance of a component in isolation,
while integrated experiments characterize the performance of a pipeline of
components, taking into account the communication and coordination overheads
of their orchestration. Experiments were performed on two Cray systems: \bw{}
at NCSA, and \titan{} at ORNL\@.

We use two metrics to characterize the performance of individual components:
throughput and concurrency. As seen in \S\ref{sec:rp}, the RP Agent is
designed as a pipeline of distinct components with multiple instances. For
each instance of a component, throughput measures the rate at which units are
managed, concurrency the volume of concurrently managed units. We measure the
throughput of a component as the number of units it handles per second,
concurrency as the number of units it handles at a given point in time.

We use two different metrics to characterize the integrated performance of
the RP Agent: total time to execution of the given workload (TTX) and
resource utilization (RU). TTX is a measure of how fast a set of tasks can be
executed by the RP Agent. It includes the time taken by the RP Agent to
manage and spawn the units for execution and the time taken by the units to
execute. RU is a measure of the percentage of available core-time spent
executing the workload and/or the RP Agent. TTX and RU are relevant for HPC
resources, which traditionally have been designed to execute large parallel
jobs and maximize overall utilization.

Depending on the type of experiment, the number of units, number or cores per
unit, duration of the unit, number of instances of a component, and number of
cores of a pilot are configurable parameters. By varying the values of these
parameters, we measure the amount of units that are in a specific state as a
function of time, or the time duration spent in a specific state. For
example, we measure the number of units in state \T{A\_SCHEDULING} and
\T{A\_EXECUTING} at every point in time in the RP Agent Scheduler component
and derive the throughput of that component.

To capture all of the measurements mentioned above, RP is instrumented with a
profiling facility to record timestamps of its operations. As the execution
of a given workload proceeds (as described in \S\ref{sub:models}), each
state transition is recoded as an event. These events are written to disk for
postmortem analysis via dedicated utility methods. RP's profiler is designed
to be non-invasive and to have minimal effect on the runtime. We measured the
temporal overhead of the profiler with a dedicated benchmark: For the same
workload executed on the same resources, the overall running time of the
Agent was \(144.7 \pm 19.2 s\) with profiling, and \(157.1 \pm 8.3 s\)
without. Note how the standard deviation of the two measurements overlap,
making the difference between the two execution times statistically
insignificant.

The execution of workloads with multiple tasks on a pilot has a varying
degree of concurrency, depending on the total number of cores required by the
tasks and available on the pilot. When the pilot has fewer available cores
than what is required by the workload, a group of tasks are executed
sequentially. We call this group of tasks a `generation'. The number of
generations of a workload execution affects the theoretical minimum TTX of
that execution. For example, given a workload with 128 single-core, 10
minutes-long tasks and a pilot with 64 cores, the execution of that workload
will require 2 generations. The theoretical minimum TTX of $2~generations
\times 10~minutes = 20~minutes$, assuming 100\% RU of the pilot's cores and
no RP Agent overhead.

It is fundamental to understand that the executable of a unit is irrelevant
to the set of experiments performed here: whether a unit runs \T{sleep},
\T{stress}, an emulator (e.g., \T{Synapse}), a simulation kernel
(e.g. \T{Gromacs}) or any other executable has no effect on the measure
of the throughput and concurrency of the RP Agent components, or on TTX and
RU\@. This follows from the design and separation of scheduling, launching
and execution of a process. The RP Agent schedules and launches a unit and,
once launched, the unit executes its code. While the unit is executing, the
Executer component of RP Agent will not interact with the unit. What code the
unit is executing is completely irrelevant to the Executer and therefore to
RP as a whole.

\subsection{Microbenchmark Experiments}\label{sub:micro}

Microbenchmarks measure the performance of individual RP components \I{in
isolation}. In a microbenchmark experiment, RP launches a pilot on a resource
with a single unit scheduled onto the Agent. When the unit enters the
component under investigation, it is cloned a specified number of
times---10000 for experiments in this paper. The components operate on the
clones, experiencing real loading while being stressed in isolation and
independent of other components.

Microbenchmark experiments are designed to isolate a component by eliminating
communication, coordination and concurrency with other components. In this
way, the benchmarked component does not compete for shared system resources
and communication channels, and remains immune from bottlenecks in other
components. Thus, the microbenchmark measures the performance \I{upper bound}
of a component implementation, as achieved \I{in isolation} from all types of
overhead as a consequence of interaction with other components.

We perform microbenchmark experiments for the Scheduler and Executer
components of RP Agent, the two components that most affect the overall
performance of the RP Agent (see
Fig.~\ref{fig:agent-arch_unit_pilot-state-models}). For the Executer, we test
two launch methods: ORTE-CLI, and ORTE-LIB\@. Note that these methods are not
used by the \textbf{executable} of the units, but instead by the RP component
to launch the executable. In turn, the executable could be
single/multi-thread/process or use MPI itself. Depending on the launch
methods, we run microbenchmarks load-balancing among 2, 4 and 8 Executer
instances, executed on 1, 2, 4, and 8 compute nodes.

We perform microbenchmark experiments on \bw{} as the representative Cray
system. As noted before, the executable of the units has no bearing on the
microbenchmarks. Microbenchmarking of the Scheduler component require no
execution, while Executer benchmarking requires actual execution of the
units. We use the \T{sleep} command to avoid any irrelevant complication
deriving from setting up specific execution environments.

A full set of microbenchmarks would span a large parameter space, making it
unfeasible to present the full set of experimental results. We focus on
results which expose performance and scaling differences among the RP Agent
components. This enables a better characterization of the overall performance
of the Agent.

\subsubsection{Agent Scheduler Performance}\label{subsec:sched_exp}

Currently, RP can instantiate exactly one Scheduler component per Agent. The
Scheduler is compute and communication bound: the scheduling algorithm
searches repeatedly through the list of managed cores, while core assignment
and deassignment are handled in separate, message-driven threads.

\begin{figure}
  \subfloat[][]{{\includegraphics[width=0.50\columnwidth]{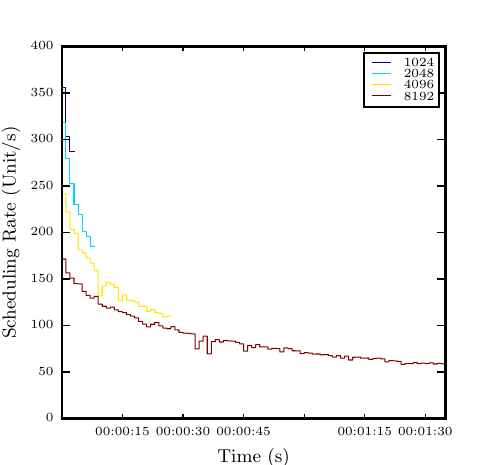}}}
  \subfloat[][]{{\includegraphics[width=0.50\columnwidth]{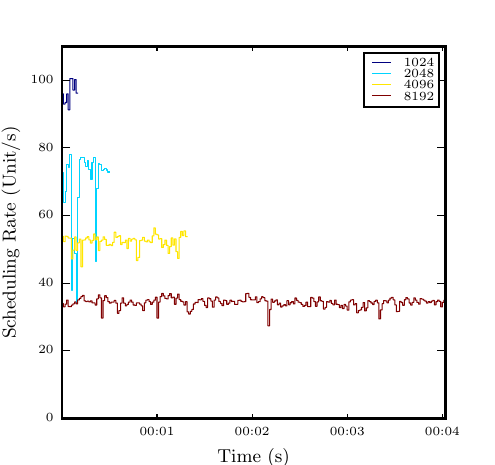}}}
  \caption{RP Agent Scheduler component throughput as function of time. 1
  generation of single-core units on 4 pilot sizes. \B{(a)} allocating cores
  to a unit; \B{(b)} both allocating cores to a unit and deallocating those
  cores from the units (steady state).}\label{fig:micro_sched}
\end{figure}

Fig.~\ref{fig:micro_sched}(a) shows the performance of the Scheduler
component in assigning cores to one generation of single-core units, for four
pilot sizes. We see that the throughput is dependent on the pilot size, and
that the throughput rate declines as more units are scheduled. This is
explained by the chosen scheduling algorithm and its implementation: the
fewer free cores remain, the more work needs to be done by the scheduling
algorithm to find a suitable set of cores for the next units. This behavior
is a consequence of using one scheduler to handle workloads with both
homogeneous and heterogeneous units (single/multi-core, mpi, cpu/gpu, etc.).
In \S\ref{sub:large_scale}, we show how a special-purpose scheduler
drastically improves performance.

Fig.~\ref{fig:micro_sched}(b) shows the same workload of the previous
microbenchmark experiment, but the measurements also include the operations
of unscheduling units and freeing cores (i.e., steady state scheduler). We do
not observe the slope of Fig.~\ref{fig:micro_sched}(a) because both the
scheduling and unscheduling operations contend the lock on the Scheduler data
structure. This considerably reduces the performance of the Scheduler when
compared to only allocating cores to the units.

\subsubsection{Unit Execution Performance}\label{subsec:uexec_exp}

RP can instantiate multiple Executer component instances per Agent. The
Executor's performance is bound by the launch methods used to spawn the units
for execution. Currently, RP supports four launch methods on Cray (ALPS, CCM,
ORTE-CLI, and ORTE-LIB). Only the two ORTE-based methods enable
single/multi-core units within and across compute nodes to run, at scales
comparable to the size of \bw{} and \titan{}.

\begin{figure}
  \subfloat[][]{{\includegraphics[width=0.50\columnwidth]{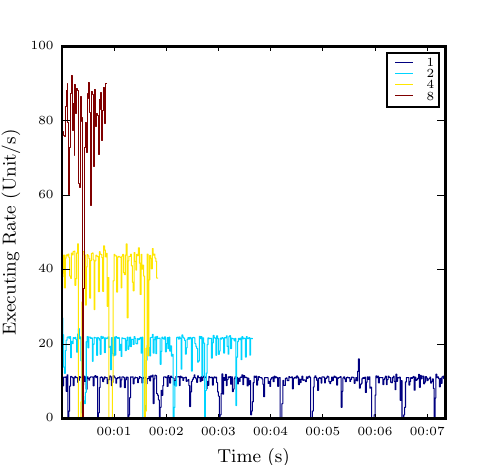}}}
  \subfloat[][]{{\includegraphics[width=0.50\columnwidth]{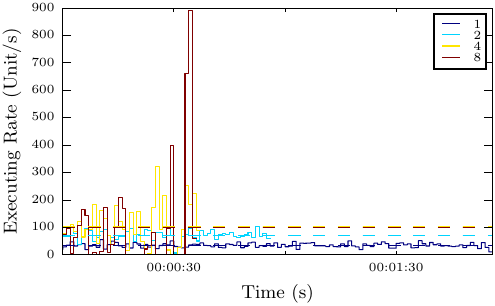}}}
  \caption{Throughput of 1--8 RP Agent Executer components with 2 launch
  mechanisms. \B{(a)} ORTE-CLI, 1--8 Executer components, each run on 1
  compute node. \B{(b)} ORTE-LIB, 1--8 instances, all run on the same MOM
  node.}\label{fig:micro_exec_orte}
\end{figure}

Fig.~\ref{fig:micro_exec_orte}~(top) shows the scaling behavior of the
ORTE-CLI launch method. Throughput scales with the number of Executer
components, with each component running on a dedicated compute node. Data for
experiments with increasing instances per node are not presented, as no
performance improvements were observed. This suggests that the current
performance of the Executer component using ORTE-CLI has an upper-bound due
to interaction with the OS\@.

While ORTE-CLI did not scale with multiple instances of an Executer component
on a single compute-node, Fig.~\ref{fig:micro_exec_orte}~(bottom) shows that
with the ORTE-LIB launch method, performance scales with up to 4 instances
per node. Adding more instances does not increase the performance further.
This suggests that 4 Executer components on 1 compute node and the ORTE-LIB
launch method reach the performance upper-bound of the ORTE layer.

\begin{figure}
  \centering
  \includegraphics[width=0.60\columnwidth]{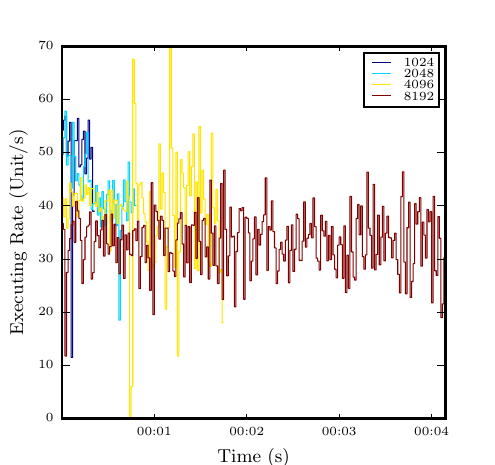}
  \caption{Throughput of 1 RP Agent Executer component with ORTE-LIB launch
  method; 1024, 2048, 4096 and 8192
  cores/units.}\label{fig:micro_exec_ortelib_cores}
\end{figure}

Fig.~\ref{fig:micro_exec_ortelib_cores} shows the scaling of the ORTE-LIB
launch method for different pilot sizes. We launch 1024, 2048, 4096 and
8192 single-core units on pilots with 1024, 2048, 4096 and 8192 cores.
Throughput is stable over time but jittery with a mean (std.\ dev) of
48.2~(10.2), 42.6~(7.1), 39.1~(9.8) unit/s. The jitter is explained by the
interaction with many external system components which, in their totality,
introduce significant noise.

The best performance of ORTE-CLI is lower than the performance of the
Scheduler for a pilot with up to 1024 cores, as seen in
Fig.~\ref{fig:micro_sched}. This indicates that ORTE-CLI creates a bottleneck
at the launching stage in the Agent's Executer. In absolute terms, the
performance of ORTE-LIB is lower than the scheduling component's when the
pilot size is less than 8192 cores, and comparable (or at times higher) at
pilot sizes over 8192 cores. Similar to the Scheduler component, the
performance decreases with increased pilot size, from an average rate of
around 48 units/s for the 1024 pilot size to an average rate of around 33
units/s for larger pilots.

\newpage
\subsection{Agent Integrated Performance}\label{subsec:agent_exp}

To characterize the RP Agent performance as a whole, we employ workloads with
varying unit durations executed on pilots of different sizes. The size of
each unit is set to 1 core, allowing experiments to measure the performance
of RP with maximum pilot cores/unit ratio. Workloads with multi-core units
lower the overall stress on the components of the Agents and their
communication and coordination protocols, resulting in better performance.

Microbenchmarks are not sufficient to characterize the Agent performance as a
whole for three reasons: (i) by definition, the microbenchmarks in
\S\ref{subsec:uexec_exp} and \S\ref{subsec:sched_exp} cannot measure the
performance cost of communication among components; (ii) the concurrent
operation of multiple components introduces competition for shared system
resources (e.g., competing for filesystem access); and (iii) the Agent
performance can be limited by components or system resources \I{outside} of
the Agent (e.g., RP client manager components, or network latency between the
Agent and MongoDB).

Accordingly, the set of integration experiments discussed in this subsection
investigates the contributions of communication and concurrency to the Agent
performance. To offset external overheads, we design the experiments so that
the Agent operates independent of the performance of the PilotManager and
UnitManager components (Fig.\ref{fig:agent-arch_unit_pilot-state-models}): we
introduce a startup barrier in the Agent to ensure that the Agent receives
sufficient work to fully utilize the pilot's resources. In this way, the
Agent starts to process units only when the complete workload has arrived at
the Agent.

On \bw{}, we measure time-dependent concurrency achieved by the RP Agent for
pilots with 2048, 4096, 8192, and 16384 cores. For each pilot size, the
workload is comprised of 3 generations of single-core units, resulting in
workloads with 6144, 12288, 24576, and 49152 units. For each workload,
the duration of each unit is 64, 128 and 256 seconds, long enough for all
the units of the first generation to start before the first unit is
completed. In this way, the first generation can always reach maximum
concurrency, saturating the number of cores available on the pilot.

\begin{figure}
  \centering
  \includegraphics[width=0.80\columnwidth]{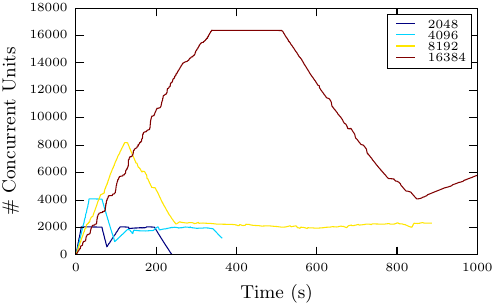}
  \caption{Unit concurrency as a function of pilot size and unit
  duration.}\label{fig:concurrency_ttx_1}
\end{figure}

Fig.~\ref{fig:concurrency_ttx_1} shows the maximal concurrency for each pilot
size, where all cores are simultaneous used to execute units. The initial
slope up to that maximum concurrency is determined by the performance of the
scheduler, which, as shown in Fig.~\ref{fig:micro_sched}(a), is dependent on
the pilot size.  For example, with the 8192-core pilot we see that 8192
units are started in about 100 seconds. This is comparable to what is shown
in figure Fig.~\ref{fig:micro_sched}(a), where 8192 units are scheduled in
about 90 seconds, with a throughput which starts out at 150 units/seconds and
later stabilizes at about 50 units/second.

Fig.~\ref{fig:concurrency_ttx_1} shows also that once the first generation of
units begins to finish execution, the scheduler enters a different mode of
operation where scheduling and unscheduling threads compete (see discussion
of Fig.~\ref{fig:micro_sched}(b)). This decreases the overall throughput of
the Agent which is no longer able to maintain full concurrency. This effect
is independent of pilot size and number of units.

\subsubsection{Comparing ORTE to ALPS and CCM}

One of the limitations of ALPS/APRUN is that we can only run one unit per
node. SSH based launch methods in CCM-mode on \bw{} are also limited, due to
connection limits when executing more than 8 concurrent units per node.  ORTE
does not have that limitation.  In order to keep the runs comparable, i.e.,
to execute the same configurations for all experiments, we configure the
workload used to use 32 cores per unit, so that each unit consumes a full
node. This workload can be executed with all RP launch methods.

On \bw{}, we run 10 workloads ranging from 3 to 768 units, where each unit
consumes a full compute node (32 cores) and executes on pilots ranging form
32 cores (1 node) to 8192 cores (256 nodes) respectively. We run the same
set of 10 workloads for each launch method and compare the actual Time to
Execution (TTX) against the theoretically optimal TTX (i.e., the time taken
by all the units to execute without any RP overhead).

\begin{figure}  
  \centering
  \includegraphics[width=0.80\columnwidth]{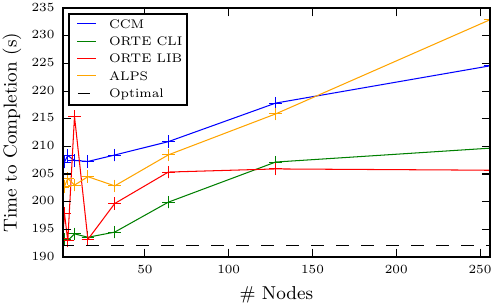}
  \caption{Time to Execution (TTX) as a function of number of units, size of
  pilot and Executer launch method.}\label{fig:concurrency_ttx_2}
\end{figure}

Fig.~\ref{fig:concurrency_ttx_2} shows that there is a large trend difference
between ORTE-CLI/ORTE-LIB and ALPS/CCM\@. As the scale increases, the
difference between ALPS/CCM to ORTE increases, with ORTE being closer to the
theoretically optimal TTX and ALPS/CCM increasing somewhat linearly after
around 50 nodes.

\subsection{Resource Utilization and Overheads at
Scale}\label{sub:large_scale}

Currently, the ORTE launch method is the one supporting the largest runs with
RP on Cray machines, allowing to execute workloads with 16384 multi-core
units on more than 130000 cores. We run two experiments to characterize the
weak and strong scaling behavior of RP and its overheads up to this scale. In
the weak scaling experiment, we execute workloads with a constant ratio
between units and cores. In the strong scaling experiment, we execute one
workload on a progressively larger amount of cores. In this way, the strong
scaling experiment executes the workload with between 2 and 32 generations.

Weak and strong scaling experiments execute workloads with 32 cores, 15
minutes long MPI tasks. We perform both experiments on \titan{} for two main
reasons: (i) \titan{} is very similar to \bw{} in terms of architecture and
scale; and (ii) these experiments required around 25 million core-hours, at
the time available only on \titan{}.

\begin{figure}
  \centering
  \includegraphics[width=0.90\columnwidth]{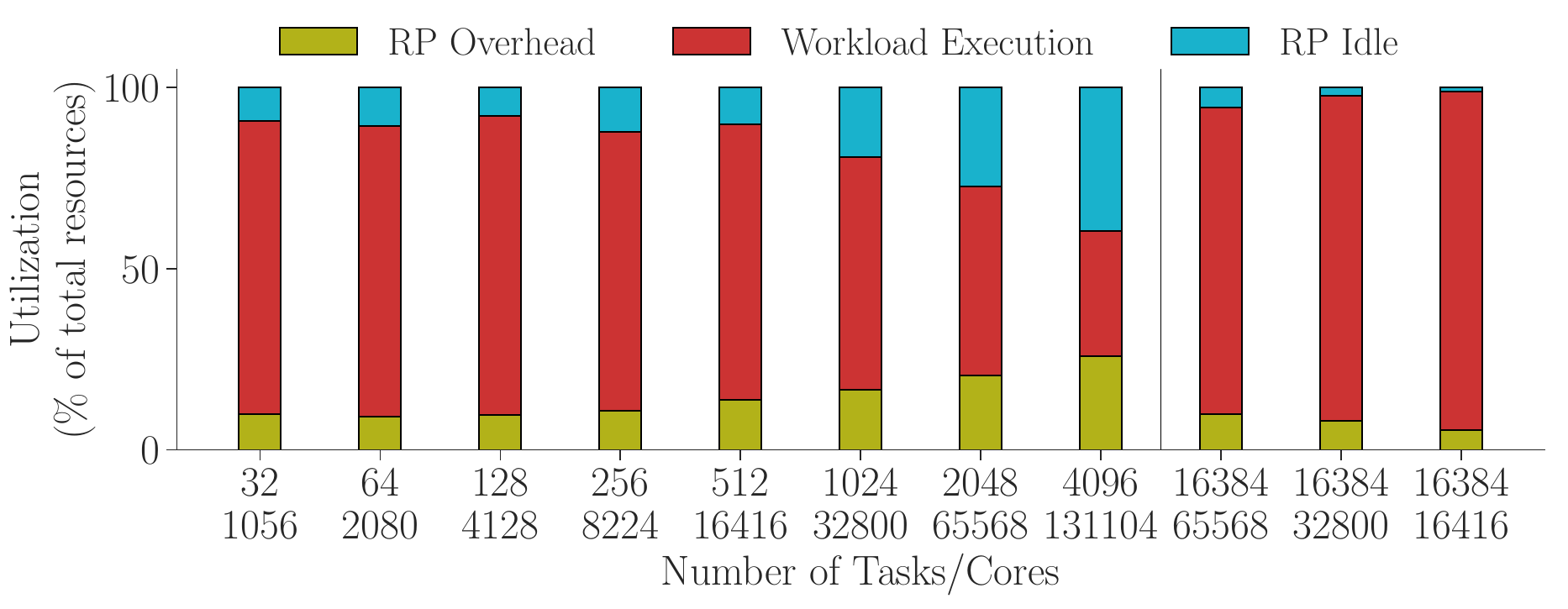}
  \caption{Resource utilization of RADICAL-Pilot.}\label{fig:ws-ru-util}
\end{figure}

Fig.~\ref{fig:ws-ru-util} shows both the weak (first 8 bars) and strong (last
3 bars) scaling experiments. We measure resource utilization as percentage of
the available core-time spent executing the workload (Workload Execution,
central portion of the stacked bar), RP code (RP Overhead, lower portion of
the stacked bar), or idling (RP Idle, top portion of the stacked bar). Runs
measuring weak scaling with between 32/1024 and 128/4096 tasks/cores have a
relatively constant percentage of core-time utilization but this percentage
decreases with the growing of the number of tasks/cores. As a result, we
observe that RP Agent does not weak scale with pilot larger than 8192 cores.

Runs measuring strong scaling show values of RP overhead and idling inversely
proportional to the number of generations: the more generations, the less
overhead and idling. This is explained by noting that, when tasks of one
generation terminate, those of the following generation immediately starts
executing. This eliminates the idling of cores for all generations but the
last one. We presume that the increase of RP overhead depends, at least to
some extent, on the proportional relation between the communication required
to coordinate an execution and the size of the pilot used.

\subsubsection{Reducing RP overhead}\label{sub:reduce_overhead}

We explore the decrease in resource utilization measured in the weak scaling
experiment (Fig.~\ref{fig:ws-ru-util}, first 8 bars) by looking at the
results of the microbenchmark shown in Fig.~\ref{fig:micro_sched}, and
focusing on the relation between scheduling performance and size of
the pilot used for the execution.

As described in \S\ref{subsec:sched_exp}, the larger the pilot, the
larger is the resource pool managed by the scheduler. Currently, the
scheduler is implemented to repeatedly search a Python data structure for
available cores. This approach is effective for a general purpose scheduler
that needs to handle many types of workload---e.g.,
homogeneous/heterogeneous, MPI/OpenMP/Scalar, or single-node/multi-node.
However, for homogeneous workloads, a more efficient single-purpose scheduler
can be implemented.

\begin{figure}
  \centering
  \includegraphics[width=0.90\columnwidth]{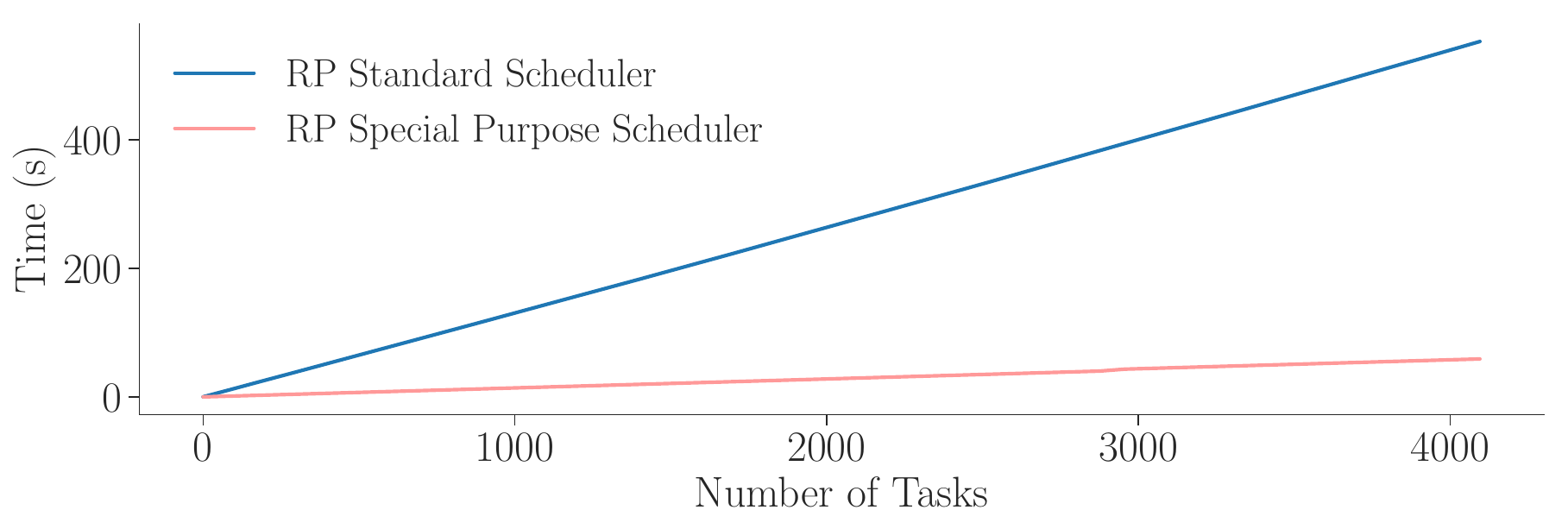}
  \caption{Scheduling overheads: Standard and special purpose
  schedulers.}\label{fig:ws-ru-sched}
\end{figure}

Leveraging the flexibility and extensibility of RP (as also used for the
Executer and its multiple launch methods), we implemented a scheduler
algorithm which specifically handles homogeneous, multi-node tasks of 
workloads used in weak and strong scaling experiments. The behavior of
this special purpose scheduler is shown in Fig.~\ref{fig:ws-ru-sched}
the scheduler manages each task in constant time, at a much lower time per
task compared to the general-purpose scheduler.

When the special purpose scheduler encounters the first unit to schedule, it
immediately divides the total set of cores into partitions which are all of
the same size as the number of cores required by the first unit. In this way,
the scheduling algorithm is reduced to the procedure of assigning
equally-sized partitions to the units as they arrive. Crucially, this avoid
the need for any search on a (Python) data structure representing the cores
managed by the pilot. Instead, partition lookup and assignment can be performed in constant time. 

It should be noted that there still remain limitations for when the second
generations of units gets scheduled, i.e., when the scheduling and
unscheduling processes compete. Nonetheless, the throughput of this scheduler
is consistently higher than for the general-purpose scheduler: the lock
contention reduces due to the reduced time for which the scheduler algorithm
needs to lock the data structures. Full details on the homogeneous bag of
task scheduler and more detailed measurements are discussed
in~\cite{rp_titan}.

\subsection{Discussion}\label{sec:discuss}

Microbenchmark experiments provide insight on how the Agent's Scheduler and
Executer components perform for different Agent configurations and pilot
sizes (\S\ref{sub:micro}). These experiments provide an upper-bound of the
throughput (i.e., units handled per second) of each of the two components and
show which component could be the rate-determining factor in the overall
agent integrated performance scales. Microbenchmark experiments for the
Agent's Scheduler component show that the scheduling throughput is dependent
on the pilot size, and that the throughput rate declines as more units are
scheduled. Further, we show that when the component is doing both the
scheduling and unscheduling operations (i.e., scheduler reaches steady state)
the scheduler's throughput is primarily dependent on the pilot size as seen
in Fig.~\ref{fig:micro_sched}(b), viz., as the pilot size increases, the
scheduler's throughput decreases.

The microbenchmark experiments with the Executer component show that its
throughput decreases when the pilot size and unit counts increase in the same
proportion, similar to what was observed for the Scheduler component. For
both ORTE-CLI and ORTE-LIB Executer launch methods we show increased
throughput when an increasing number of concurrent executers are used.
ORTE-LIB allows multiple executers on the MOM node, while ORTE-CLI requires a
compute node for each executer. As described in \S\ref{sub:orte-arch}, this
is explained by observing that an execution through ORTE-LIB is only a
library call that causes a network call and doesn't strain the system on
which it is running. Conversely, each execution call done through ORTE-CLI
requires interactions with the filesystem and network resources to
communicate with the \T{orte-dvm}. Thus an agent using ORTE-CLI reaches the
resource limits of \bw{} and Titan with workloads that consists of very large
numbers of concurrent tasks or when running multiple components on the same
MOM node.

\S\ref{subsec:agent_exp} describes experiments that characterize the
integrated performance of RP Agent. We show that the maximal concurrency
achievable for multiple pilot sizes, where all cores are simultaneously used
to execute units, is approximately 16000 units. We also compare the
performance of the ALPS, CCM and ORTE Executer launch methods and found that
ORTE-LIB and ORTE-CLI launch methods out-perform ALPS and CCM for the
performance metric of TTX\@. We also show that the performance of ORTE-CLI
and ORTE-LIB launch methods are comparable when the number of units is
between 3 and 768 units, but this is likely to change with a higher number of
units, where configurations with multiple executers and the lower impact of
ORTE-LIB on resource utilization would make the ORTE-LIB launch method
perform better than ORTE-CLI\@. Further, we note that the performance of CCM
on \bw{} is low compared to other launch methods available on non-Cray HPC
systems like Stampede~\cite{review_radicalpilot}.

Finally, we measured the resource utilization of RP Agent at highest scale
currently supported, both in terms of number of units concurrently executed
and number of cores of a single pilot. We moved from a single-core units to
units of 32-cores, and we ran weak scaling experiments with workloads ranging
from 32 to 4096 units on pilot sizes ranging from 32 to 131104 cores
respectively, and strong scaling experiments with workloads of 16384 units
on pilot sizes ranging from 16416 to 65568 cores (\S\ref{sub:large_scale}).
Our experiments show that resource utilization of the RP Agent for the weak
scaling experiments with pilot sizes between of 1024 and 4096 cores have a
relatively constant percentage of core-time utilization, but this percentage
significantly decreases with pilots larger than 8192 cores. We attributed RP
Agent's poor weak scaling property with pilot sizes over 8192 cores to the
performance of the Agent Scheduler component.

We addressed the decrease in resource utilization measured in the weak
scaling experiment and showed the flexibility and extensibility or RP, by
implementing a special-purpose scheduler, specific to the experimental
workload---i.e., homogeneous, multi-node tasks. We then showed that the
overhead added by the special-purpose scheduler significantly decreases
compared to the one of the special-purpose scheduler used for the experiment.

\section{Conclusion}\label{sec:conclusion}

{\it Prima facie}, a system implementing the Pilot
abstraction~\cite{review_pilotreview,pstar12} provides the conceptual and
functional capabilities needed to support the scalable execution of many task
workloads. The impact of an abstraction is limited to its best
implementation. Whereas there are several existent pilot systems, they are
either geared towards specific functionality or platforms. This paper
describes the architecture and implementation of RP (\S\ref{ssec:arch}), and
characterizes the performance of its Agent module on Cray platforms
(\S\ref{sec:exp}).

In molecular sciences, there is a demonstrated need~\cite{nextgen-molsim} to
be able to support up to 10$^5$ MPI tasks as part of a single ``ensemble
simulation''. Similar scales are anticipated across multiple domains. Several
parts of RP will need to be re-engineered to efficiently execute workloads at
this scale. Most of the benefits will come from improving the Agent, as
discussed in \S\ref{ssec:arch} and consistent with results shown in
\S\ref{sec:exp}. To this end, we are planning to: (i) develop a set of
specialized, lock-free schedulers; (ii) partition the pilot resources and
operate multiple agents in parallel on these partitions; (iii) explore new
launch methods; and (iv) aggregate units depending on their application
provenance and duration to optimize Scheduler and Executor throughput.

The focus of this paper has been on the direct execution of workloads on HPC
machines, but RP also serves as the runtime system for a range of other tools
and libraries~\cite{entk,repex2016,power-of-many17,extasy}, many already used
in production. The requirements of these tools and libraries will also
motivate future research and development.\\

\section*{Software and Data} 

RP is available for immediate use on many platforms~\cite{radical_pilot_url}.
RP source is accompanied with extensive documentation and an active
developer-user community. Source code, raw data and analysis scripts to
reproduce experiments can be found at:

\begin{itemize}
  \item RADICAL-Pilot: {\scriptsize \url{https://github.com/radical-cybertools/radical.pilot}}
  \item RADICAL-Analytics: {\scriptsize \url{https://github.com/radical-cybertools/radical.analytics}}
  \item Data and scripts: {\scriptsize \url{https://github.com/radical-experiments/jsspp18}}
\end{itemize}

\section*{Acknowledgments}

This work is supported by NSF ``CAREER'' ACI-1253644, NSF ACI-1440677
``RADICAL-Cybertools'' and DOE Award DE-SC0016280. We acknowledge access to
computational facilities: XSEDE resources (TG-MCB090174) and Blue Waters
(NSF-1713749).

\bibliographystyle{splncs}
\bibliography{rp-jsspp2018}

\begin{thebibliography}{10}

\bibitem{better-resource}
Hwang, E., Kim, S., k.~Yoo, T., Kim, J.S., Hwang, S., r.~Choi, Y.:
\newblock Resource allocation policies for loosely coupled applications in
  heterogeneous computing systems.
\newblock IEEE Transactions on Parallel and Distributed Systems \textbf{PP}(99)
  (2015)

\bibitem{review_pilotreview}
Turilli, M., Santcroos, M., Jha, S.:
\newblock A comprehensive perspective on {Pilot-Jobs}.
\newblock {ACM Computing Surveys (accepted, in press)} (2017) \newline
  \url{http://arxiv.org/abs/1508.04180}.

\bibitem{preto2014fast}
Preto, J., Clementi, C.:
\newblock Fast recovery of free energy landscapes via diffusion-map-directed
  molecular dynamics.
\newblock Physical Chemistry Chemical Physics \textbf{16}(36) (2014)
  19181--19191

\bibitem{cheatham2015impact}
Cheatham~III, T.E., Roe, D.R.:
\newblock The impact of heterogeneous computing on workflows for biomolecular
  simulation and analysis.
\newblock Computing in Science \& Engineering \textbf{17}(2) (2015)  30--39

\bibitem{sugita1999replica}
Sugita, Y., Okamoto, Y.:
\newblock Replica-exchange molecular dynamics method for protein folding.
\newblock Chemical physics letters \textbf{314}(1) (1999)  141--151

\bibitem{pordes2007open}
Pordes, R.,  et~al.:
\newblock {The Open Science Grid}.
\newblock J. Phys.: Conf. Ser. \textbf{78}(1) (2007)

\bibitem{maeno2014evolution}
Maeno, T., De, K., Klimentov, A., Nilsson, P., Oleynik, D., Panitkin, S.,
  Petrosyan, A., Schovancova, J., Vaniachine, A., Wenaus, T.,  et~al.:
\newblock Evolution of the {ATLAS} {PanDA} workload management system for
  exascale computational science.
\newblock In: Proceedings of the 20th International Conference on Computing in
  High Energy and Nuclear Physics (CHEP2013), Journal of Physics: Conference
  Series. Volume 513(3)., {IOP} Publishing (2014)  032062

\bibitem{raicu2007falkon}
Raicu, I., Zhao, Y., Dumitrescu, C., Foster, I., Wilde, M.:
\newblock Falkon: a {Fa}st and {L}ight-weight tas{K} executi{ON} framework.
\newblock In: Proceedings of the 8th {ACM}/{IEEE} conference on Supercomputing,
  {ACM} (2007) ~43

\bibitem{wilde2011swift}
Wilde, M., Hategan, M., Wozniak, J.M., Clifford, B., Katz, D.S., Foster, I.:
\newblock Swift: A language for distributed parallel scripting.
\newblock Parallel Computing \textbf{37}(9) (2011)  633--652

\bibitem{CCM}
{CCM}:
\newblock \footnotesize\url{http://bit.ly/cray_ccm} (accessed January 2018).

\bibitem{karo2006application}
Karo, M., Lagerstrom, R., Kohnke, M., Albing, C.:
\newblock The application level placement scheduler.
\newblock (2006)

\bibitem{castain:orte2007}
Castain, R.H., Squyres, J.M.:
\newblock {Creating a transparent, distributed, and resilient computing
  environment: the OpenRTE project}.
\newblock The Journal of Supercomputing \textbf{42}(1) (October 2007)  107--123

\bibitem{taskfarmer}
{TaskFarmer}:
\newblock \footnotesize\url{http://bit.ly/taskfarmer}.

\bibitem{wraprun}
{Wraprun}:
\newblock \footnotesize\url{https://www.olcf.ornl.gov/kb_articles/wraprun/}.

\bibitem{qdo}
{QDO}:
\newblock \footnotesize\url{http://bit.ly/nersc_qdo}.

\bibitem{canon2012my}
Canon, R.S., Ramakrishnan, L., Srinivasan, J.:
\newblock My {Cray} can do that? {S}upporting diverse workloads on the {Cray
  XE-6}.
\newblock Cray User Group (2012)

\bibitem{pythontaskfarm}
{Python Task Farm}:
\newblock
  \newline\footnotesize\url{http://www.archer.ac.uk/documentation/user-guide/batch.php#sec-5.7}.

\bibitem{ahn2014flux}
Ahn, D.H., Garlick, J., Grondona, M., Lipari, D., Springmeyer, B., Schulz, M.:
\newblock Flux: A next-generation resource management framework for large hpc
  centers.
\newblock In: Parallel Processing Workshops (ICCPW), 2014 43rd International
  Conference on, IEEE (2014)  9--17

\bibitem{saga-x}
Merzky, A., Weidner, O., Jha, S.:
\newblock {SAGA}: A standardized access layer to heterogeneous distributed
  computing infrastructure.
\newblock Software-X (2015) DOI: 10.1016/j.softx.2015.03.001.

\bibitem{cug-2016}
Santcroos, M., Castain, R., Merzky, A., Bethune, I., Jha, S.:
\newblock Executing dynamic heterogeneous workloads on blue waters with
  radical-pilot.
\newblock In: Cray User Group 2016. (2016)

\bibitem{pmix}
{PMIx web site}:
\newblock \footnotesize\url{https://www.open-mpi.org/projects/pmix/}.

\bibitem{cffi}
{CFFI Documentation}:
\newblock \footnotesize\url{http://cffi.readthedocs.org}.

\bibitem{rp_titan}
Merzky, A., Turilli, M., Maldonado, M., Jha, S.:
\newblock Design and performance characterization of {RADICAL}-pilot on titan.
\newblock arXiv preprint arXiv:1801.01843 (2018)

\bibitem{review_radicalpilot}
Merzky, A., Santcroos, M., Turilli, M., Jha, S.:
\newblock {Executing Dynamic and Heterogeneous Workloads on Super Computers}
  (2016) (under
  review)\newline\footnotesize\url{http://arxiv.org/abs/1512.08194}.

\bibitem{pstar12}
Luckow, A., Santcroos, M., Merzky, A., Weidner, O., Mantha, P., Jha, S.:
\newblock P*: A model of pilot-abstractions.
\newblock IEEE 8th International Conference on e-Science (2012)  1--10
  ~~\footnotesize \url{http://dx.doi.org/10.1109/eScience.2012.6404423}.

\bibitem{nextgen-molsim}
{Shantenu Jha and Peter M. Kasson}:
\newblock {High-level software frameworks to surmount the challenge of 100x
  scaling for biomolecul\ ar simulation science}.
\newblock {White Paper submitted to NIH-NSF Request for Information (2015)}
  \newline\footnotesize\url{http://dx.doi.org/10.5281/zenodo.44377}.

\bibitem{entk}
Balasubramanian, V., Treikalis, A., Weidner, O., Jha, S.:
\newblock Ensemble toolkit: Scalable and flexible execution of ensembles of
  tasks.
\newblock In: 2016 45th International Conference on Parallel Processing (ICPP).
  Volume~00. (Aug. 2016)  458--463

\bibitem{repex2016}
Treikalis, A., Merzky, A., Chen, H., Lee, T.S., York, D.M., Jha, S.:
\newblock Repex: A flexible framework for scalable replica exchange molecular
  dynamics simulations.
\newblock In: 2016 45th International Conference on Parallel Processing (ICPP).
  (Aug 2016)

\bibitem{power-of-many17}
Balasubramanian, V., Turilli, M., Hu, W., Lefebvre, M., Lei, W., Cervone, G.,
  Tromp, J., Jha, S.:
\newblock {Harnessing the Power of Many: Extensible Toolkit for Scalable
  Ensemble Applications}.
\newblock (2017) ~~\footnotesize \url{https://arxiv.org/abs/1710.08491}.

\bibitem{extasy}
Balasubramanian, V., Bethune, I., Shkurti, A., Breitmoser, E., Hruska, E.,
  Clementi, C., Laughton, C., Jha, S.:
\newblock Extasy: Scalable and flexible coupling of md simulations and advanced
  sampling techniques.
\newblock In: 2016 IEEE 12th International Conference on e-Science (e-Science).
  (Oct 2016)  361--370

\bibitem{radical_pilot_url}
{RADICAL-Pilot}:
\newblock \footnotesize
  \url{https://github.com/radical-cybertools/radical.pilot}.

\end{thebibliography}

\end{document}